\documentclass{an-think}

\usepackage{graphicx,fancyhdr}
\usepackage{times}

\begin{document}

% -------------- Title and affiliation
%                (please edit accordingly)

\title{MASTER: The Mobile Astronomical System of Telescope-Robots.}

\author{V.M.Lipunov\inst{1,}\inst{2}\fnmsep \thanks{lipunov@sai.msu.ru,
http://observ.pereplet.ru},\noindent A.V.Krylov\inst{1},%\fnmsep\thanks{krylov@sai.msu.ru},
V.G.Kornilov\inst{1},%\fnmsep\thanks{victor@sai.msu.ru}
G.V.Borisov\inst{1},%\fnmsep\thanks{borisov@sai.crimea.ua},
D.A.Kuvshinov\inst{1},\\%\fnmsep\thanks{dmitriyk@genphys.phys.msu.su},
A.A.Belinsky\inst{1},%\fnmsep\thanks{aleks@sai.msu.ru},
M.V.Kuznetsov\inst{1},%\fnmsep\thanks{mihailo@sai.msu.ru},
S.A.Potanin\inst{1},%\fnmsep\thanks{potanin@sai.msu.ru},
G.A.Antipov\inst{2},%\fnmsep\thanks{antipov@sai.msu.ru},
N.V.Tyurina\inst{1},\\%\fnmsep\thanks{tyurina@sai.msu.ru}
E.S.Gorbovskoy\inst{2},%\fnmsep\thanks{gorbovskoy@sai.msu.ru}
I.Chilingaryan\inst{2}}

\institute{Sternberg Astronomical Institute, Moscow State University, Universitetskij pr-t, 119992, Moscow, Russia
 \and Physical Department of Moscow State University, Vorob'evy Gory, 1, 119992, Moscow, Russia}

\date{Received; accepted; published online}

\abstract{We present the first russian robot-telescope designed to
make prompt observations of gamma-ray bursts
(http://observ.pereplet.ru). The telescopes are near Moscow. The
system of telescopes with prompt pointing rates connects to the
internet. The main parameters are the following: Richter-Slefogt
system telescope (355 mm, f/d=2.4); Richter-Slefogt system
telescope (200mm, f/d=2.4); Flugge system telescope (280mm,
f/d=2.5); TV-camera with 20x40 degree objective; Two CCD cameras
(Pictor 416); One CCD Apogee Camera AP16E. The type of mount is
German with 8 grad/sec slew rate. MASTER  images stars down to 19
magnitude in a 1 min exposure covering 6 square degrees.
  } \maketitle
% -----------------------------------------------------------------

\section{Introduction}

The creation of robotic observatories  is a leap forward for
modern astronomy. These complexes are especially indispensable in
survey work, being devoted to the discovery of new and known
transient phenomena, like gamma-ray bursts, supernova and nova
bursts, microlens events, comets, asteroids, etc. The position and
time of these phenomena can't be predicted. So observations from
several tens of longitudes (which Russia has) can increase the
effectiveness of new astronomical object detections.

%the goal to cover the latitudes and longitudes in the mostly in
%longitudes country can essentially raise the efficient of
%discovery and observations of new astronomical objects.

The MASTER system %$(http://observ.pereplet.ru)$
 is located 30km from Moscow. It is the first robotic telescope of its type and unique  in Russia.
  It was built to provide the rapid response necessary in the follow-up observations of gamma-ray bursts (GRBs).
   Multicolor photometry of GRB optical emission in the first
minutes after their detection gives us an opportunity to  study
the  physical processes taking place at the beginning stage of
these most powerful  radiation events in the Universe. We should
point out that up to now optical afterglows have only been
observed after delays of tens of seconds. There are no  light
curves (LC), taken in the first seconds and  extending to several
hours, taken by one instrument and  in one photometric system. The
shape of the LC will permit us to study the dynamics of
relativistic sphere expansion and to draw conclusions about the
nature of GRBs.

The second goal of the MASTER project relies on the fact  that the
optical GRB afterglow can be more isotropic then gamma-ray one and
consequently can be detected more often, since the polar pattern
in such different wavelengthes can not concur. Optical emission
can have more isotropic distribution. MASTER allows us to make a
constant sky survey in order to discover optical bursts without
gamma-ray radiation, but produced by the same physical process. We
 can survey 50 square degrees per hour to a $19.5^m$ limiting magnitude.
%higher of$19^m$ %of %than %%$19^m$.

Our third goal is  to make wide-field observations synchronously
with gamma-ray telescopes (for example, HETE).

In summary, MASTER can carry out
 \textit{early afterglow research into GRB, triggered by alert from gamma-ray telescopes},
 \textit{make synchronous observations of the region of a GRB} and \textit{ make sky surveys}.

Let's consider the main principles of the work of such robotic systems.

\section{Robot telescopes}

A robotic telescope is an autonomous instrument, that can both
point to, and make observations of a sky region by itself, then
automatically reduce the images, perform image quality checks, and
point to discovered objects. Such telescope, designed to detect
the optical radiation from a gamma-ray burst's source, works by
the following scheme:

\noindent \fbox{\parbox[b][1\height]{3.3cm}{Gamma-ray bursts Center Network(GCN) ~ sends an
Alert}}$\Rightarrow$ \fbox{\parbox[b][1\height]{2.7cm}{Robot-telescope points to received
coordinates}} $\Rightarrow$ \fbox{\parbox[b][1\height]{1.1cm}{GRB observations}}

$\Rightarrow$\fbox{\parbox[b][1\height]{2.5cm}{Image reduction
}}$\Rightarrow$\fbox{\parbox[b][1\height]{3.9cm}{GCN circular publication }}

The publication in GCN circulars is the next step in minimising the error-box boundaries for the
newly-discovered object \footnote{http://gcn.gsfc.nasa.gov/gcn3$\_$archive.html}, that makes them
observable by large telescopes %(becouse the new error-box is now enough small).

%The efficiency of telescopes-robots was demonstrated in the programs of SuperNova, asteroids and comets search.
% These systems are indispensable for synchronous and follow-up to the gamma-burst moment.

There are  about 20 autonomous observatories, MASTER included,
directed to GRB observations in optical wavelength,
%like%ROTSE, NEAT, KAIT, REM, MERCATOR and other
registered in GCN %\footnote{(http://gcn.gsfc.nasa.gov/gcn\_sites.html)}.

%\textbf{GROCSE (LLNL \& U.Mich.)} http://www-phys.llnl.gov/V\_Div/GROCSE
%\textbf{LOTIS (LLNL \& Clemson)} http://compton.as.arizona.edu/LOTIS
%\textbf{ROTSE} http://www.umich.edu/~rotse/index.html
%\textbf{TAROT} http://www.cesr.fr/~boer/tarot
%\textbf{LOTOSS/KAIT} http://astron.berkeley.edu/~bait/kait.html
%\textbf{ETC (MIT)} http://space.mit.edu/ETC/ NEAT http://neat.jpl.nasa.gov/
%\textbf{NEAT} http://neat.jpl.nasa.gov/
%\textbf{The ~BRADFORD ~ telescope ~(~U.of Bradford, England)}  http://www.telescope.org/index.php
%\textbf{Wyoming IR Obs (S.Howell)} http://physics.uwyo.edu/~mpierce/WIRO/
%\textbf{LONEOS} http://www.lowell.edu/users/elgb/loneos\_disc.html
%\textbf{Beijing Xinglong Observatory} http://www.bao.ac.cn/bao/station/xl/index-e.html
%\textbf{BOOTES (Spain, Castro-Tirado)} http://www.laeff.esa.es/BOOTES/
%\textbf{APT (Australia)} http://www.phys.unsw.edu.au/~mcba/apt.html
%\textbf{NYTT (Finland)} http://www.phys.unsw.edu.au/~mcba/apt.html
%\textbf{ASAS} http://archive.princeton.edu/~asas/
%\textbf{MERCATOR (La Palma)} http://www.mercator.iac.es/
%\textbf{REM (Italy)} http://golem.merate.mi.astro.it/projects/rem/
%\textbf{BART (Czech Rep.)} http://lascaux.asu.cas.cz/en/
%\textbf{Grove Creek Obs (Australia)}  http://www.gco.org.au/
%\textbf{ Liverpool Telescope (La Palma)} http://telescope.livjm.ac.uk/
%

The ROTSE-III (Ackerlof, 2000) instruments  are the most similar to MASTER. %ROTSE is mounted in USA, Australia, South Africa.
ROTSE-III has made follow-up observations of about 5 bursts (from
60 up to 500 sec) since the first observations of synchronous
optical afterglow of a GRB in 1998. We should point out a feature
of MASTER and ROTSE-III. These systems are able to make a rapid
response to GRB alerts in addition to an ongoing survey of the
whole available sky. The major feature is the depth of the
surveys: MASTER can measure $19^m$ objects with a 1.5 min
exposure. Furthermore  MASTER has a wide-field camera, that allows
us to make  observations synchronously with gamma-ray telescopes
in the magnitudes range $9^m$ %(when this camera staysindependently) - $13^m$(when camera is on the telescope focus) ingood weather.

\section{The MASTER main parameters}% and work principles.  }

The MASTER system includes 3 wide-aperture telescopes. These
 are
mounted on a robotic German parallax mount, which can slew at a
rate of 8$ \circ/sec$. The main MASTER Richter-Slefogt telescope
(diameter = 355mm, D:F = 1:2.4) can obtain  unfiltered images of
$19^m$ objects over 6 square degrees in a 1.5 min co-added
exposition (3 exp. of 30sec). Additionally, there is a modified
Richter-Slefogt telescope (200mm, D:F = 1:2.5) and a Flugge system
telescope (280mm, D:F = 1:2.8). All three telescopes are mounted
parallel to one another on the same mount, which allows us to make
a simultaneous  observations in R and V filters. There are 2 CCD
cameras: an AP16E camera with Kodak KAF-16801E CCD (4096 x 4096
pixels, front illuminated, 30 sec download time) on the  355mm
telescope, and a Pictor-416 (700 x 500 pixels) on the 200mm one.

There is also a wide field video camera (fov = 30x40 deg). This
camera allows us to work synchronously with space-based gamma-ray
telescopes (being pointed to the region of the GRB), giving a
magnitude limit of $8^m$. This was used for making a film of an
aurora \footnote{http://observ.pereplet.ru/indexr.shtml}, which
was observed over
Moscow on 2003/10/30. %One can see a thin structure of this phenomena in these frames.

MASTER is permanently connected to the internet and receives alerts from GCN.

MASTER works in the following way:

\fbox{\parbox[b][1\height]{1.1cm}{manager
server}}$\Rightarrow$\fbox{\parbox[b][1\height]{3.4cm}{programm block(PB) of telescope management}}
$\Rightarrow$ \fbox{\parbox[b][1\height]{1.5cm}{PB to get images}}

$\Rightarrow$\fbox{initial image reduction PB} $\Rightarrow$\fbox{the database}.

To obtained a fast and precise telescopes pointing, two identical
drives were mounted on the mount, one on the declination axis, and
one on the  polar axes. Each  drive consists of a step-motor and
intelligent controller, connected to manager computer through the
RS485 interface.
%Controller provides the travel to set distance with set velocity
%and acceleration and braking parameters.

The managing program (run on a Linux operating system) executes
telescope pointing commands from the managing server. The program
converts from astronomical object coordinates to the instrumental
coordinate system. It also calculates the optimal route of
telescope displacement (if the object is available at the current
time) keeping within the limits accessible to the telescope.
% The last circumstance is especially actual, because of the German mount, so as pointing to chance
%coordinates needs the telescope transposition in $30\%$ cases.

The rough calibration  of the instrumental coordinates system is
supplied by zero-points sensors. This approximate correction is
applied to objects with known coordinates. This simple system
supplies a pointing with a precision of 5", which is adequate
given the instruments field of view.

\section{MASTER optical observations of GRB}

MASTER began surveying for  GRB optical afterglows  at the end of
2002. Our main results are summarized in Table 1. We should note
that no  GRB with an optical afterglow occurred during local
nighttime from Dec 2002  to May 2004.

We publish only the first observation of the GRB error-box (if
anyone observes the alert error-box before us and there was no
optical afterglow, we do not observe it again). GRB030329
excludes, we present  only upper limits on the magnitude (i.e.
there is no afterglow to these mag). The main time delay is the
time required for data reduction and transmission by the gamma-ray
telescope (HETE in our case). The MASTER telescopes can slew into
position in 10-15sec.

\begin{table*}[] \centering \small
  \caption{ Observations of GRB error-boxes.}\label{Tab1}
\begin{tabular}{|c|c|c|c|c|}
\hline GRB     &  Number of GCN       & Time $t-t_0$ from& Optical mag & Comments\\
&circular publication& GRB alert &image limits & to observations\\

\hline GRB040308 & GCN 2262&  48h &  $21.2$& Unfiltered\\

\hline GRB030913& GCN 2394  & 43 min& $17.5$ & Clouds present for 120 seconds after GRB alert .\\
& GCN 2385&&&The first image was taken only 43 minutes after  GRB alert.\\
%There were clouds in 120 seconds after GRB time .\\
% So the first image was only 43 minutes later than GRB time.\\
%OT limit - 17.5 unfiltered instrumental magnitude.\\

\hline GRB030601 & GCN 2262& 55 min&$12.0$&Pointing in 105 seconds. There're clouds first 55min \\
%& & &&There were clouds in first 55 min\\
 %seconds after trigger time. Clouds.  First measurable image of the error box area was taken 55 minutes after the burst. OT limit 12.0.\\

\hline GRB030414& GCN 2157&8h&$13.3 - 14.5$&First observations. Almost full Moon, clouds.\\
%First observation . 8 hours after the GRB. OT limit 13.3 - 14.5. Background level was high due to cirrus\\
% clouds illuminated by Moon (almost full).\\

\hline GRB040308 &GCN 2543&48h&$21.2$&First observations\\

\hline GRB030418&GCN 2158&11h&$16.5$&First observations\\

\hline GRB030416 &GCN 2154&33h&$15.0$&Full Moon. First observations\\

\hline GRB030328 & GCN 2103& 5h&$18.3$&First observations\\

\hline GRB030329 &GCN 2091, GCN 2035   &5.2h&&First observations  in Europe.\\
& GCN 2002&&& R-filter brightness curve for 8.8h\\

\hline GRB021219 &GCN 1770& 7.5h&$13.7$&First observations of first INTEGRAL GRB, clouds. \\
%&&&& INTEGRAL GRB, clouds.\\

% 5,5 hours after GRB time (first observation in Europe), R-light curve during 7 hours.\\

 \hline

\end{tabular} \normalsize  \medskip
\end{table*}

The scientific mean of the limits on found  optical flux can be estimated in the following way. All
the published  GCN observations of GRB optical afterglows  (unfiltered and R) are in the Figure 1.
Stellar magnitudes limits of the observed error-boxes were reduced to the gamma wavelength fluence
of GRB030329% by the following
: $m = m_{obs} + 2.5 log (F_\gamma / F_{030329})$. The dashed-lines at the Figure 1 are the bounds
of normal fluences. If the light of the optical afterglow is less than
low limit, we'll be able to detect "dark"-GRB. %There
%are our upper limits of GRB021219, GRB030414, GRB030416 in this region. %The first point is by Akerlof(1999) at the
%burst-time.

\section{GRB030329 observation.}

Images of optical transient GRB030329 (Peterson,Price,2003) were
taken by MASTER (Lipunov et al.,GCN2002, GCN2035) on the
2003/03/29. We note that MASTER was the first instrument in Europe
to start observations of this event. Figure 2 presents the optical
light curve for this GRB event
\footnote{http://observ.pereplet.ru/lightcurve.txt}, consisting of
more than 200 frames. To increase the  signal/noise ratio and to
reduce instrumental errors we averaged ? times of exposure span,
resulting in 64 data points. The magnitude calibration was
performed in accordance with GCN2023 (Henden,2003) using 4 stars.
The large errors on some of the  magnitudes are due to weather
conditions; the first data points were obtained during sunset with
high sky illumination, while the last data points were affected by
cirrus cloud.

An analysis  of our light curve gives the following approximation during 8.8h of observations
($t-t_0$ in days from 5.2h to 14.0h)

$F \sim t^{-\alpha},$ ~~~~~ $\alpha=1.22 \pm 0.03$.

$R_{mag} = 15.8 + 1.2*2.5*log(t - t_0)$

The power index of our light curve corresponds to the observations
made by Burenin et al., 2003 within error-limits.

\begin{figure}[t]
%   \centering
   \includegraphics[angle=270,width=9cm]{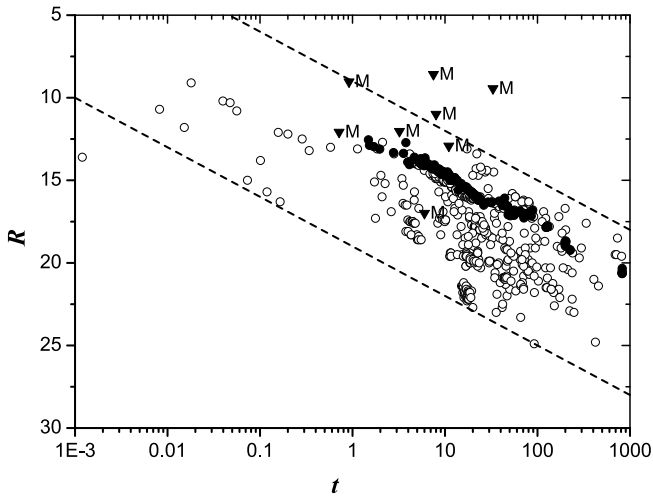}
   \caption{"Synthetic" brightness curve. Y-axis is R or unfiltered magnitude, X-axis is the time from the burst moment in hours.
   Triangles are the optical limits, found by MASTER.
   The dashed-line inclination corresponds to the low of light of
   optical afterglow of GRB030329, $t^{-1.2}$. The circular are the observational data of almost all GCN circulars.}
\end{figure} \begin{figure}[t]
   \centering  \includegraphics[angle=90,width=8.5cm]{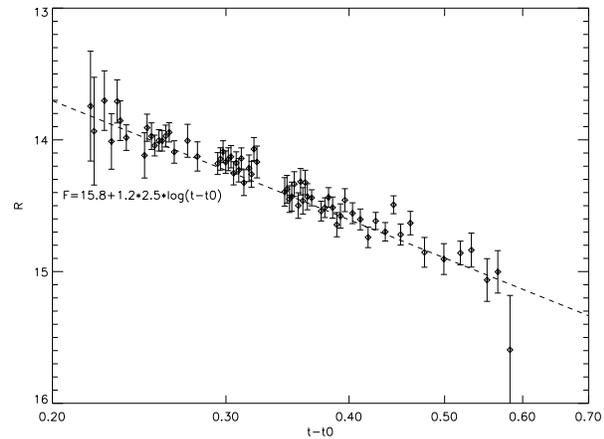}
   \caption{Brightness curve of GRB030329 measured by MASTER. Y-axis is R-magnitude, X-axis is the time from the burst in days.}
   \label{F1}
\end{figure}

%\textit{ Here we represent pre-final light curve of the OT. We used more than 200 direct images to obtain field
%photometry. Several groups of images were summarized to achieve better S/N ratio and smaller uncertanities of
%measurements. As the final result, we represent 64 points. Magnitude calibration was made according to the data,
%presented in GCN 2023 (A. Henden). 4 stars were used to do the calibration. Big uncertainties of magnitude
%measurements are caused by weather conditions. First points were made just after sunset on the blue sky, last points
%were made through cirruses near the horizont. From these data we obtained power law flux decreasing during 8 hours of
%observations ($t-t_0$ from 5.2h to 13.5h): New light curve plots available at old locations:http...}GRB030329_R_lc

\section{Conclusions}

The MASTER system of robotic telescopes was designed to make
follow up and synchronous observations of  gamma-ray burst events.
Such systems allow us to study the earliest stages of GRB optical
emission and also allow us to find supernova and  new transient
optical phenomena in the Universe. At the same time, we can
discover minor planets, comets and space garbage. It's especially
important to address this problem using small telescopes since
observations can only be made by large telescopes some hours after
the alert.

We present our light curve of GRB030329 taken in the R-filter.

Synthetic light curve of GRB  observations is introduced. One can estimate received optical limits
on observed error-boxes of GRBs. There are regions for normal GRB and for "dark" GRB.

 \acknowledgements Authors are deeply indebted to S.M.Bodrov, A.V.Bagrov, E.Yu.Osminkin, T.D.Krylova, D.K.Magnitsky. This work is partly supported by RFFI
04-02-16411.

\end{document}